\documentclass{iau}
\usepackage{graphicx}

\title[Clustering of Radiogalaxies in the Las Campanas Redshift Survey] 
{Problems  of  Clustering of Radiogalaxies}

\author[W{\l}odzimierz God{\l}owski, Agnieszka Pollo \& Jacek Golbiak]   
{W{\l}odzimierz God{\l}owski $^1$,
Agnieszka Pollo $^2$,$^3$
 \and  Jacek Golbiak $^4$}

\affiliation{$^1$ , Institute of Physics, Opole University, Oleska 48, 45-052 Opole, Poland 
 \\ email: {\tt godlowski@uni.opole.pl } \\[\affilskip]
$^2$ Astronomical Observatory, Jagellonian University,   Orla 171, 30-244 Krakow, Poland \\[\affilskip]
$^3$ National Centre for Nuclear Research, Hoza 69, 00-689 Warszawa
 \\email: {\tt apollo@camk.edu.pl} \\[\affilskip]
$^4$  Institute of Philosophy of Nature and Natural Sciences, John Paul II Catholic University of Lublin;,
Al. Rac³awickie 14, 20-950 Lublin, Poland
\\email: {\tt jgolbiak@kul.lublin.pl}}

\pubyear{2012}
\volume{290}  
\jname{Feeding compact objects: Accretion on all scales}
\editors{C.M. Zhang, T. Belloni, M. M\'endez \& S.N. Zhang, eds.}

\begin{document}

\maketitle

\begin{abstract}

We present the preliminary analysis of clustering  of a sample of 1157
radio-identified galaxies from \cite{a3}. We found that for separations
$2-15 h^{-1}$Mpc their redshift space autocorrelation function 
$\xi(s)$ can be approximated 
by the power law  with the correlation length $\sim 3.75h^{-1}$Mpc and 
slope $\gamma \sim 1.8$.
The correlation length for radiogalaxies is found to be lower and the 
slope steeper than the corresponding parameters of the control sample 
of optically observed galaxies. 
Analysis the projected correlation function $\Xi(r)$ displays possible 
differences in the clustering  properties between active 
galactic nuclei
(AGN) and starburst (SB) galaxies. 

\end{abstract}

\begin{figure}[b]
\begin{center}
 \includegraphics[width=3.5in,height=1.2in]{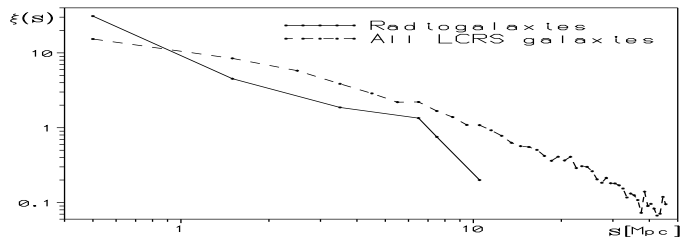}
 \caption{Redshift space autocorrelation function $\xi(r)$ 
for optical galaxies and radiogalaxies (for historical reasons $h$=0.5 is assumed).}
   \label{fig1}
\end{center}
\end{figure}

\begin{figure}[b]
\begin{center}
 \includegraphics[width=2.2in,height=1.00in]{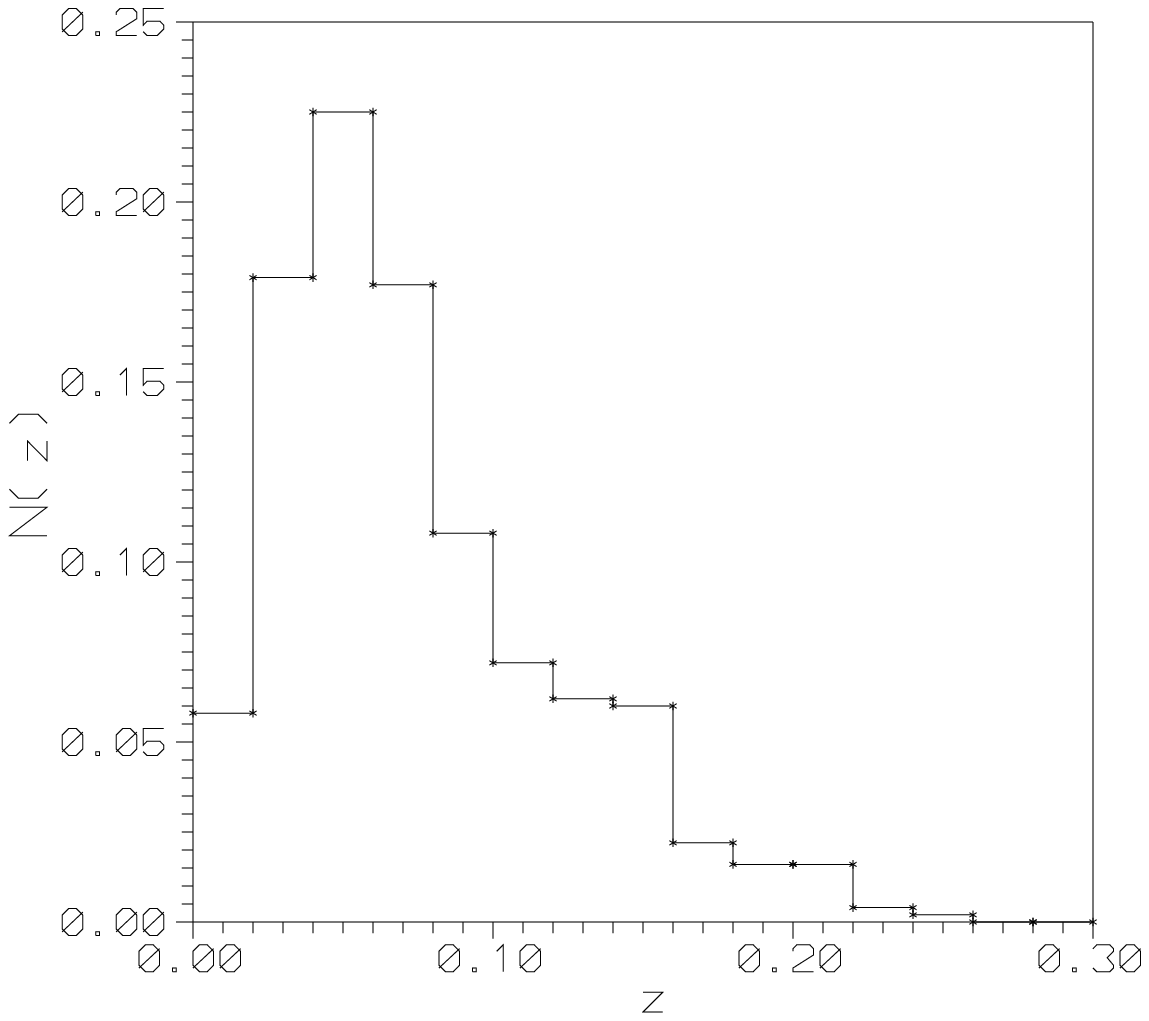}
 \includegraphics[width=2.2in,height=1.00in]{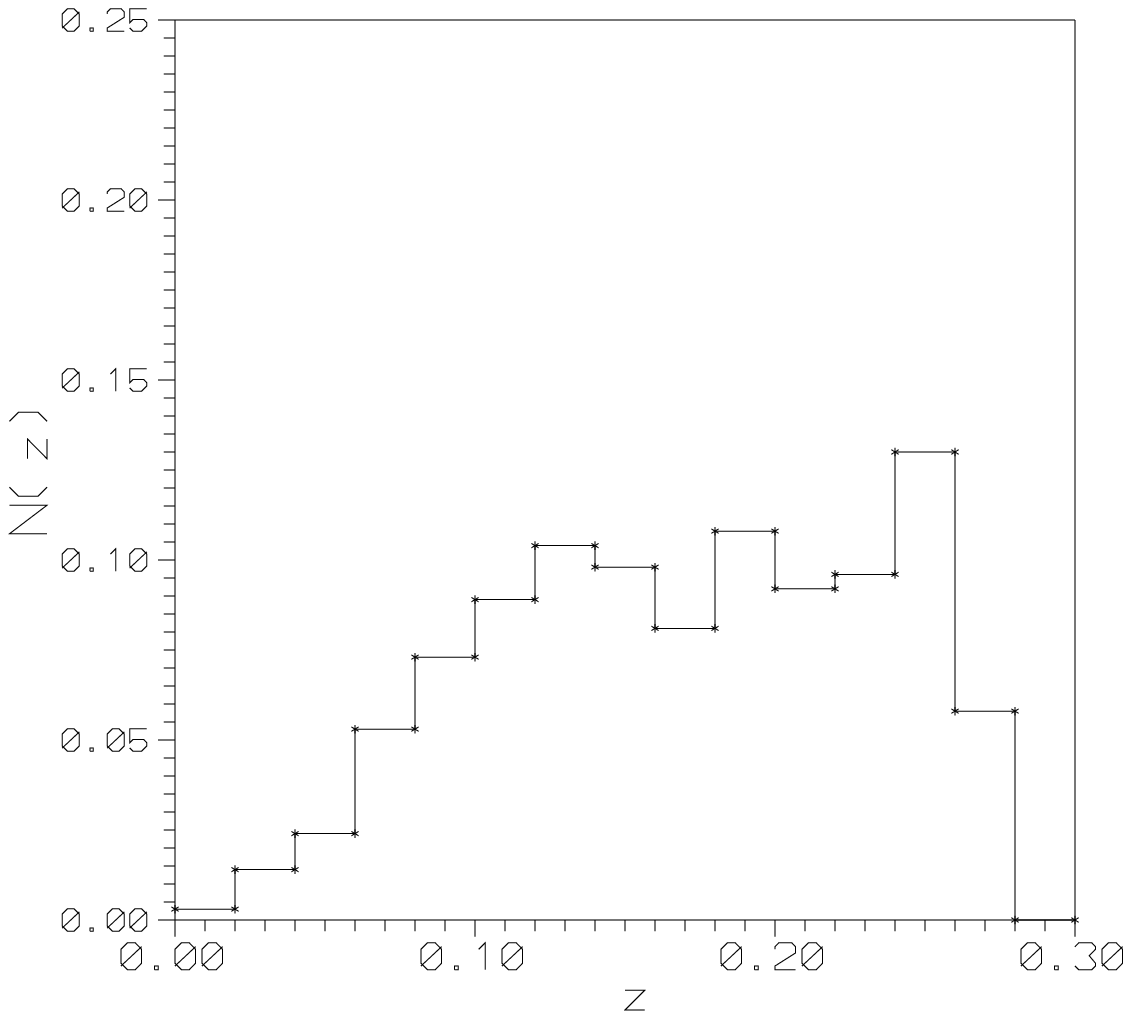}
 \caption{Redshift distribution of radiogalaxies: SB (left panel) and AGN (right panel).}
   \label{fig2}
\end{center}
\end{figure}

Clustering of the radiogalaxies  was first detected by  \cite{a4} in the redshift
survey of 329 galaxies with $z<0.1$ and radio fluxes $S(1.4$GHz$)>500$mJy (hereafter PN91).
They found that the redshift space correlation function of these radiogalaxies
 could be fitted by $\zeta=[s/11h^{-1}$Mpc$]^{-1.8}$ where $s$ is the galaxy-galaxy 
separation in the redshift space. Later \cite{a5} analyzed the sample of 451
radio  identified galaxies selected from the LCRS 
(\cite{a6}) and NVSS (\cite{a1}) surveys. Using the projected correlation function
$\Xi(r)=\int\xi[(r^2 +x^2 )^{0.5}]dx$ he found that for the optical galaxies
the correlation length was $\sim 5h^{-1}Mpc$ while the correlation length 
for the radio-loud subsample was $\sim 6.5h^{-1}Mpc$. \cite{a5} suggested
that these differences resulted from the fact that his sample was dominated
by starburst (SB) galaxies while the majority of the PN91 sample were luminous 
active galactic nuclei (AGNs). In the present paper we analyze clustering 
of radiogalaxies using 
the sample of 1157 galaxies selected from the LCRS and NVSS by \cite{a3}.

We measured angular and spatial autocorrelation functions $w(\theta)$ and
$\xi(s)$ both for optical galaxies and radiogalaxies 
The measurement for optical galaxies remains very similar to that obtained 
by \cite{a7}. The angular correlation function of radiogalaxies is characterized 
by the slope $\epsilon=\gamma-1=0.97 \pm0.10$ for ALL the sample, with 
$\epsilon=1.13 \pm0.14$ for AGNs and $\epsilon=0.86\pm0.14$ for SB galaxies. 
The $w(\theta)$ deviates from the  power law on a scale $0.37^o$, which 
is lower than the value obtained for optical LCRS  galaxies ($0.54^o$).
The redshift space 
correlation function for radiogalaxies can also be approximated by a power law; the 
measured correlation lengths are $3.75 h^{-1}$Mpc$\pm0.4$ for all classes of 
radio sources (ALL, AGN and SB). However, slopes are different 
in all cases: we obtain the value of $\gamma=1.76\pm0.11$ for a general sample 
of radiogalaxies, $2.39\pm0.34$  for AGNs and  $1.66\pm0.15$ for SB galaxies  
The redshift space correlation function displays the same feature as the 
angular correlation function: a different slope for optical and radiogalaxies, 
and inside a radio-loud sample - a different slope for AGNs and SB galaxies. 
In general,  the value of $\gamma$ for radiogalaxies is higher than that found 
for optical galaxies: $\gamma=1.52 \pm0.04$ (see Fig. 1). The analysis of the 
projected autocorrelation  function $\Xi(r)$  suggests that the correlaton 
length is higher for AGNs than for SB galaxies. In the same time, the measured 
correlation length for AGNs is lower than that obtained for optical LCRS 
galaxies. In our opinion, the differences between all the samples are a mixed 
effect of two factors: a selection bias (mainly different redshift 
distributions for AGNs and SB galaxies, see Fig. 2) and different 
environments of AGN and SB galaxies.

Our main  result is that for separations between $2-15h^{-1}$Mpc autocorrelation 
function $\xi(s)$ for radiogalaxies can be approximated by the power law 
with slope $\gamma \sim 1.8$ and correlation length $\sim 3.75h^{-1}$Mpc.
The measurements of the projected autocorrelation function $\Xi$ suggest that 
the correlation length for AGNs is  higher than that for SB galaxies.
All approaches to the measurement of the correlation function show differences 
in the slope coefficients between AGN and SB radiogalaxies. 
This result can be  interpreted  as following: AGN are radio sources of
type FRI and FRII. We have no information whether a particular AGN belongs to 
the type FRI  or  FRII. However, it is clear that a significant number of the 
radio sources classified as AGNs is connected to elliptical galaxies located in
the  centers of galaxy clusters. In contrast, SB are mostly spiral galaxies,
and they are located more often in the  outer parts  of clusters or even on the 
borders of filaments where processes of galaxy  formation related to the  
starburst processes are stronger in the local Universe.

\end{document}